\documentclass[5p, preprint]{elsarticle}
\biboptions{sort&compress}
\usepackage{lineno,hyperref,isotope,mathrsfs}
\modulolinenumbers[5]
\usepackage{amsmath,bbold,amssymb,epsfig,bm,mathrsfs}
\usepackage{feynmp,color,slashed,nicefrac,exscale,multirow}
\usepackage{times,txfonts}
\begin{document}
\begin{frontmatter}
\title{$^{48}$Si: An atypical nucleus?}
\author[b]{Jia Jie Li}
\author[a]{Wen Hui Long}
\author[c,d]{J\'{e}r\^{o}me Margueron}
\author[e]{Nguyen Van Giai}
\address[a]{School of Nuclear Science and Technology, Lanzhou University, Lanzhou 730000, China}
\address[b]{Institut f\"{u}r Theoretische Physik, J. W. Goethe-Universit\"{a}t, D-60438 Frankfurt am Main, Germany}
\address[c]{Institute for Nuclear Theory, University of Washington, Seattle, Washington 98195, USA}
\address[d]{Institut de Physique Nucl\'{e}aire de Lyon, Universit\'{e} Claude Bernard Lyon 1, IN2P3-CNRS, F-69622 Villeurbanne Cedex, France}
\address[e]{Institut de Physique Nucl\'{e}aire d'Orsay, Universit\'{e} Paris-Sud, Universit\'{e} Paris-Saclay, IN2P3-CNRS, F-91406 Orsay Cedex, France}
\begin{abstract}
Based on the relativistic Hartree-Fock formalism and one of the most advanced Lagrangian PKA1, we investigate the properties of the exotic nucleus $^{48}$Si. We found that $^{48}$Si may be an atypical nucleus characterized by i) the onset of doubly magicity, ii) its location at the drip line, iii) the presence of a doubly semibubble (central depletion of the neutron and proton density profiles) in the ground state, and iv) the occurrence of pairing reentrance at finite temperature. These phenomenons are not independent from each others. We illustrate for instance that the doubly semibubble reduces the spin-orbit splitting of low-$\ell$ orbitals and modifies the splitting of relevant pseudospin partners, favoring $N=34$ as a new magic number for neutron rich nuclei. Since $^{48}$Si is predicted doubly magic, it could have an extra stability which puts it at the drip line. Moreover, $^{48}$Si may have interesting excited states which may induce pairing reentrance at finite temperature. While not being new, these phenomenons are found to serendipitously occur together in $^{48}$Si, from our theoretical calculation. Theoretical nuclear modelings are known to be poorly predictive in general, and we asset our confidence in the prediction of our modeling on the fact that the predictions of PKA1 in various regions of the nuclear chart have systematically been found correct and more specifically in the region around $^{48}$Si, our approach correctly reproduce the known features of neighboring nuclei. Whether our predictions are confirmed or not, $^{48}$Si provides a concrete benchmark for the understanding of the nature of nuclear forces.

\end{abstract}
\end{frontmatter}
%
%
Closed-shell nuclei play particularly important roles in nuclear structure physics since the ground state of nuclei having proton or neutron numbers equal to magic numbers can be considered as an archetype of independent particles moving in a spherical potential. Theoretically, they provide crucial benchmarks for mean field properties using density functional approaches~\cite{Bender2003,Meng2006,Sagawa2014}. For instance, the spin-orbit (SO) coupling~\cite{Mayer1948,Haxel1949} and the approximate pseudospin symmetry (PSS)~\cite{Arima1969,Hecht1969,Ginocchio1997} play a crucial role in the occurrence of shell closure, and are incidentally two manifestations of the relativistic nature of the nuclear interaction~\cite{Ginocchio2005,Liang2015}. The strong SO coupling existing in stable and near-stable nuclei is tightly related to the sharp surface of the nuclear potential which is self-consistently determined by the density profile~\cite{Bender2003,Todd2004,Meng2006}. The modification of the density profile, e.g. halo or central depletion, could modify the structure properties of exotic nuclei, such as their magic numbers as we will show in our present study of $^{48}$Si.

\begin{figure}[tb]
\centering
\ifpdf
\includegraphics[width = 0.48\textwidth]{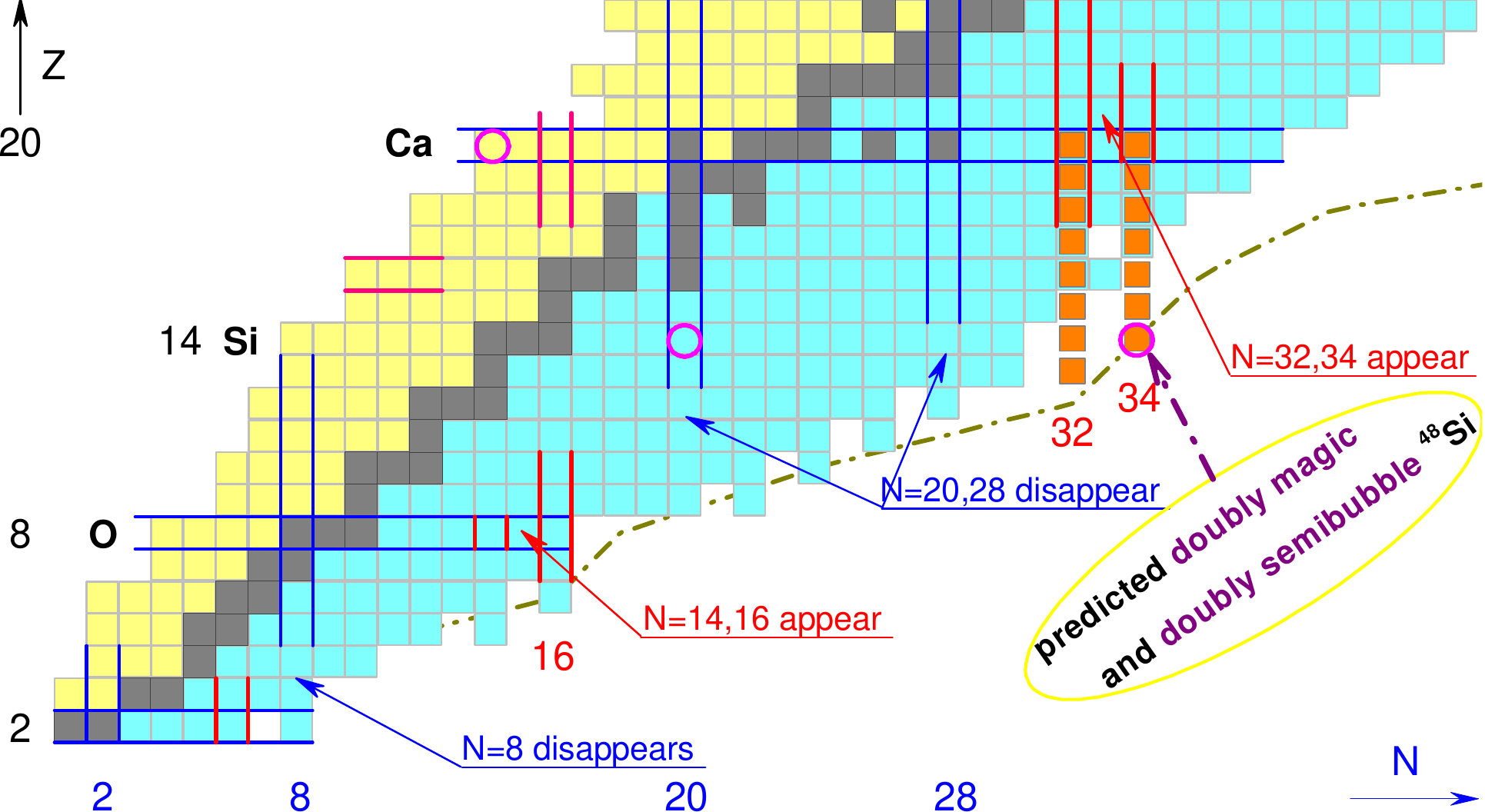}
\else
\includegraphics[width = 0.48\textwidth]{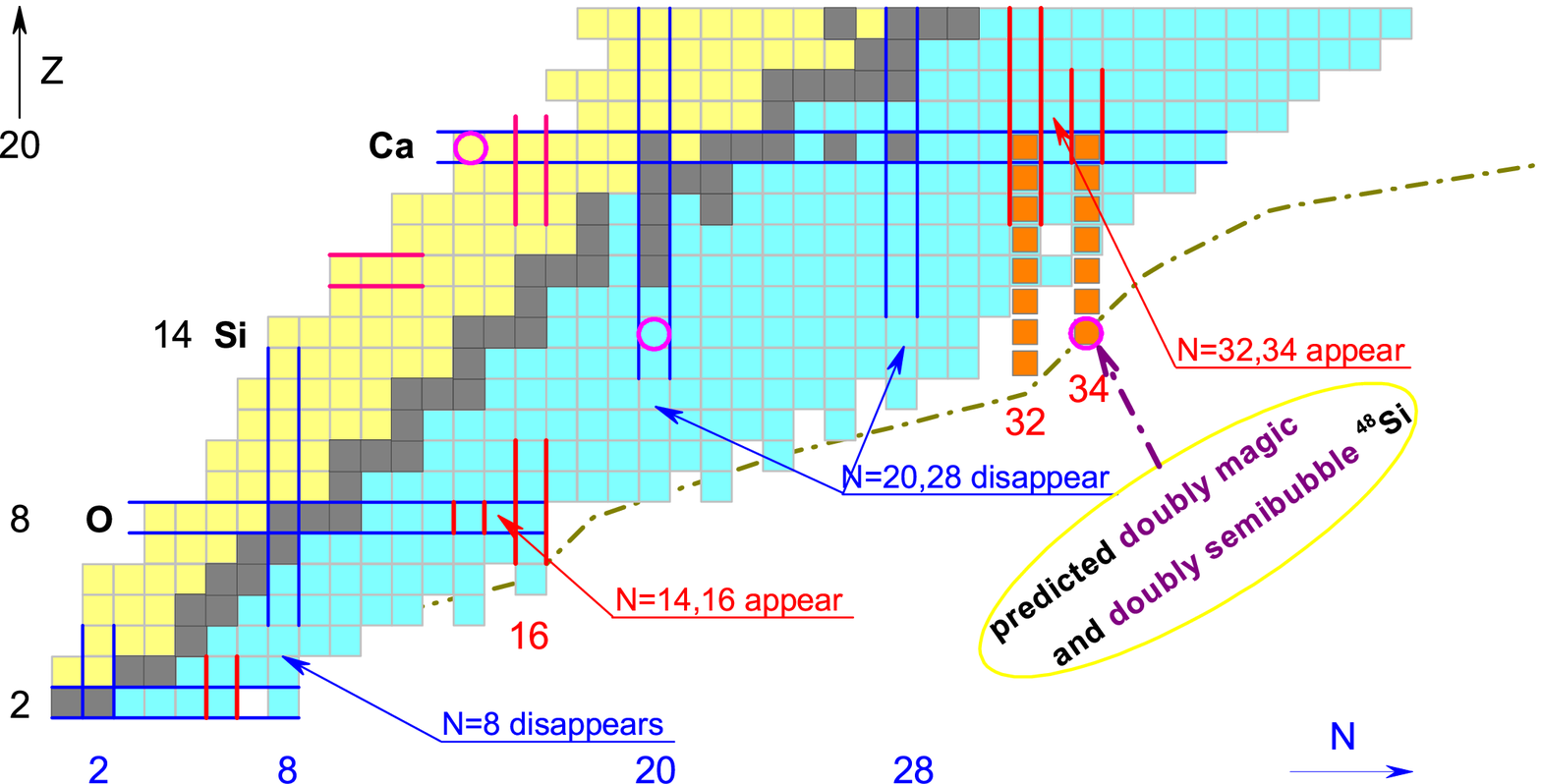}
\fi
\caption{Recent observations of the evolution of magicity, updated from Ref.~\cite{Tanihata2013}. Some isotopes of interest here are indicated in black. The green dashed-dotted line shows the average drip line predicted by relativistic energy density functionals~\cite{Afanasjev2013}. The orange squares stand for the $N=32$ and 34 isotones with $Z\leq20$. The three pink circles mark the expected bubble-like nuclei $^{34}$Ca and $^{34,48}$Si~\cite{Li2016b,Burgunder2014,Mutschler2017}.
}
\label{fig:MG}
\end{figure}

Over the past few decades significant progress in exotic nuclei has brought important changes in our view of finite nuclear systems. For instance, magic numbers are not universal across the nuclear chart and they can change dramatically depending on the number of neutrons or protons, leading to novel and unexpected features~\cite{Gade2008,Sorlin2008,Tanihata2013,Otsuka2013}. A few of them include the collapse of the conventional magic numbers $N=8$, 20 in the islands of inversion~\cite{Tanihata1985,Warburton1990,Caurier1998,Navin2000}. New magic numbers can arise, as observed in the typical case of dripline magic nucleus $^{24}$O~\cite{Ozawa2000,Hoffman2008,Kanungo2009}. Recent observations of the appearing of new closed shells in proton- and neutron-rich nuclei are summarized in Fig.~\ref{fig:MG}. These experimental achievements have demonstrated the idea that shell evolution in nuclei comes from a combination of complex effects related to the nuclear force.

More specifically in neutron-rich $pf$-shell nuclei, intensive efforts have demonstrated the occurrence of new magic shells at $N=32$ and 34. The magicity at $N=32$ have been signed from the measurements of the $2^+_1$ excitation energy in titanium, chromium and calcium isotopes~\cite{Prisciandaro2001,Dinca2005,Gade2006}, and further confirmed by the high-precision mass measurements of exotic calcium and potassium isotopes~\cite{Wienholtz2013,Rosenbusch2015}. The magicity at $N=34$ was revealed from the measurement of the $2^+_1$ energy in $^{54}$Ca~\cite{Liddick2004,Steppenbeck2013}, and theoretically supported by ab-initio calculations~\cite{Hagen2012,Hergert2014}, shell models~\cite{Steppenbeck2013,Steppenbeck2015}, and some energy density functionals containing tensor terms~\cite{Grasso2014,Yuksel2014,Li2016a}.
Very recently, the mass evolution in calcium isotopes beyond $N=34$ indicated again the magicity at $N=34$~\cite{Michimasa2018}.
Since the new magic numbers $N=32$ and 34 in calcium isotopes are now well established, it's natural to examine how they evolve in more exotic regions, e.g., in the $N=32$ and 34 isotones with less protons. Quite recently, shell model calculations have predicted a larger $N=34$ subshell gap in $^{52}$Ar than the one reported in $^{54}$Ca~\cite{Steppenbeck2015}, and an increasing of the $2^+_1$ energies in argon, sulfur and silicon isotopes from $N=32$ to 34~\cite{Utsuno2015}. It has also been shown from the relativistic approach that the $N=34$ shell gaps are continuously enhanced from $^{52}$Ar to $^{48}$Si, making $^{48}$Si a new dripline magic nucleus~\cite{Li2016a}.

The last identified silicon isotope to date is $^{44}$Si~\cite{Tarasov2007}, with four neutrons less than $^{48}$Si. The synthesis of $^{48}$Si is however a major challenge which may not be overcome in the near future. A possibility scenario for the production of more exotic silicon isotopes may be based on the new heavy-ion-induced nucleon-exchange reactions~\cite{Gade2009}, which could lead to secondary nuclei with more neutrons than the primary beam. While we measure the difficulties in the synthesis of $^{48}$Si, the purpose of the present study is to illustrate, from a theoretical viewpoint, why $^{48}$Si may be a very important nucleus for the benchmark of our present nuclear models and interactions. To our knowledge, the interest in the exploration of silicon isotopes is based on three observations: i) the existence of the new subshell closure at $Z=14$ in $^{34}$Si~\cite{Baumann1989,Caurier2014}, ii) the disappearance of the $N = 28$ shell closure in $^{42}$Si~\cite{Bastin2007,Takeuchi2012}, and iii) the possible existence of a proton semibubble (reduced density at the nuclear interior) structure  in $^{34}$Si~\cite{Grasso2009, Li2016b}, where recent experimental evidence has been reported~\cite{Burgunder2014,Mutschler2017}. Specifically, a central depletion in the nuclear density profile can modify the nuclear mean field potential in two ways, namely reducing the depth of potential well as well as the strength of SO potential in the central region of the nucleus~\cite{Todd2004,Li2016b}. In this work we present the first illustrative prediction where such a central depletion phenomenon may significantly modify the single-particle (s.p.) structure, leading eventually to the occurrence of a new magicity.

In connection to our analysis, it is interesting to mention a recent correlation analysis which has shown that the central density in medium-mass nuclei carries little information on the properties of nuclear matter since it is predominantly driven by shell structure~\cite{Schuetrumpf2017}. We can therefore deduce that for the silicon isotopes, the dominant properties are induced more by the finite size effects, e.g. surface properties, finite range and detailed properties of the nuclear interaction, rather than global properties of the nuclear interaction as it appears in the average mean field approximation for nuclear matter. In our present analysis, this remark provides an additional argument for the promotion of nuclei such as $^{48}$Si to benchmark nuclear modeling.

Let us start with the discussion of the $N=34$ and $Z=14$ new magicities as predicted from the relativistic Hartree-Fock-Bogoliubov (RHFB) approach~\cite{Long2010,Li2016a}. The coupled integro-differential RHFB equations are solved on the Dirac Woods-Saxon basis~\cite{Zhou2003}, well suited for the description on the density profile from the center to the most external part of nuclei.
We base our predictions on the PKA1 Lagrangian~\cite{Long2007} which is yet the most advanced relativistic Lagrangian, considering the number of exchanged mesons and the full and consistent treatment of Fock terms. The model Lagrangian is based on the exchange of $\pi$ (pseudovector), $\omega$ (vector), $\sigma$ (scalar) and $\rho$ (vector, tensor) mesons, as well as density dependent coupling constants. The nuclear tensor force is naturally taken into account by the Fock diagrams~\cite{Li2016a,Long2007,Jiang2015}. The $N=34$ shell gaps are predicted to be $\sim 2.5$ and 4.0~MeV, respectively, for $^{54}$Ca and $^{48}$Si~\cite{Li2016a}, being consistent with the shell model calculations~\cite{Steppenbeck2015,Utsuno2015}.
In addition, it is found that the tensor $\rho$ and pseudovector $\pi$ meson-nucleon couplings, which can be treated as a mixture
of central and tensor forces, are the important ingredients in covariant density functional to reproduce the evolutions of the s.p. spectrum for both $sd$- and $pf$-shell nuclei~\cite{Li2016a,Li2016b}.
This results exemplify that the PKA1 Lagrangian furnishes a optimal choice for discussing the properties of very neutron-rich nuclei.

\begin{figure}[b]
\ifpdf
\includegraphics[width = 0.48\textwidth]{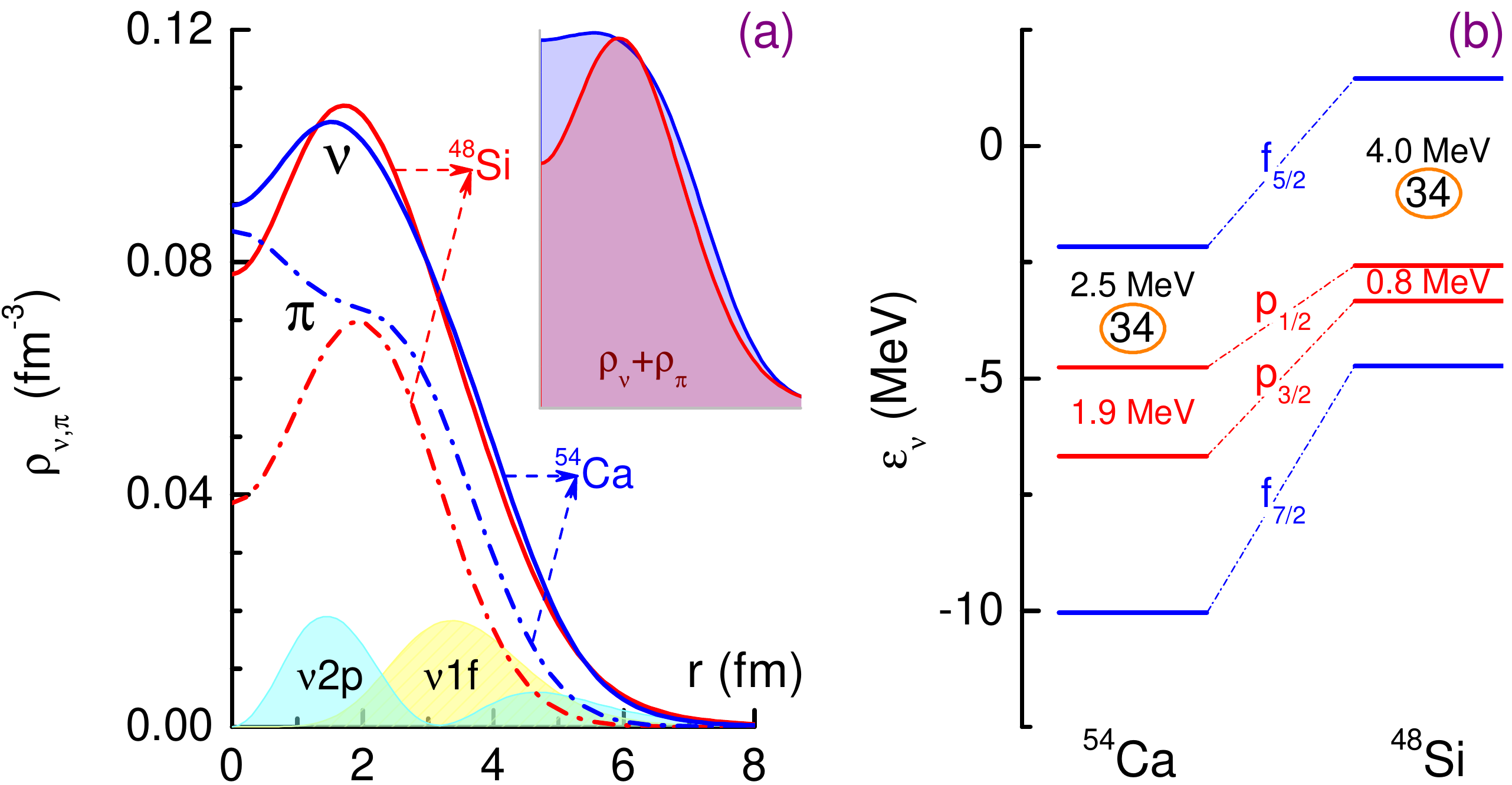}
\else
\includegraphics[width = 0.48\textwidth]{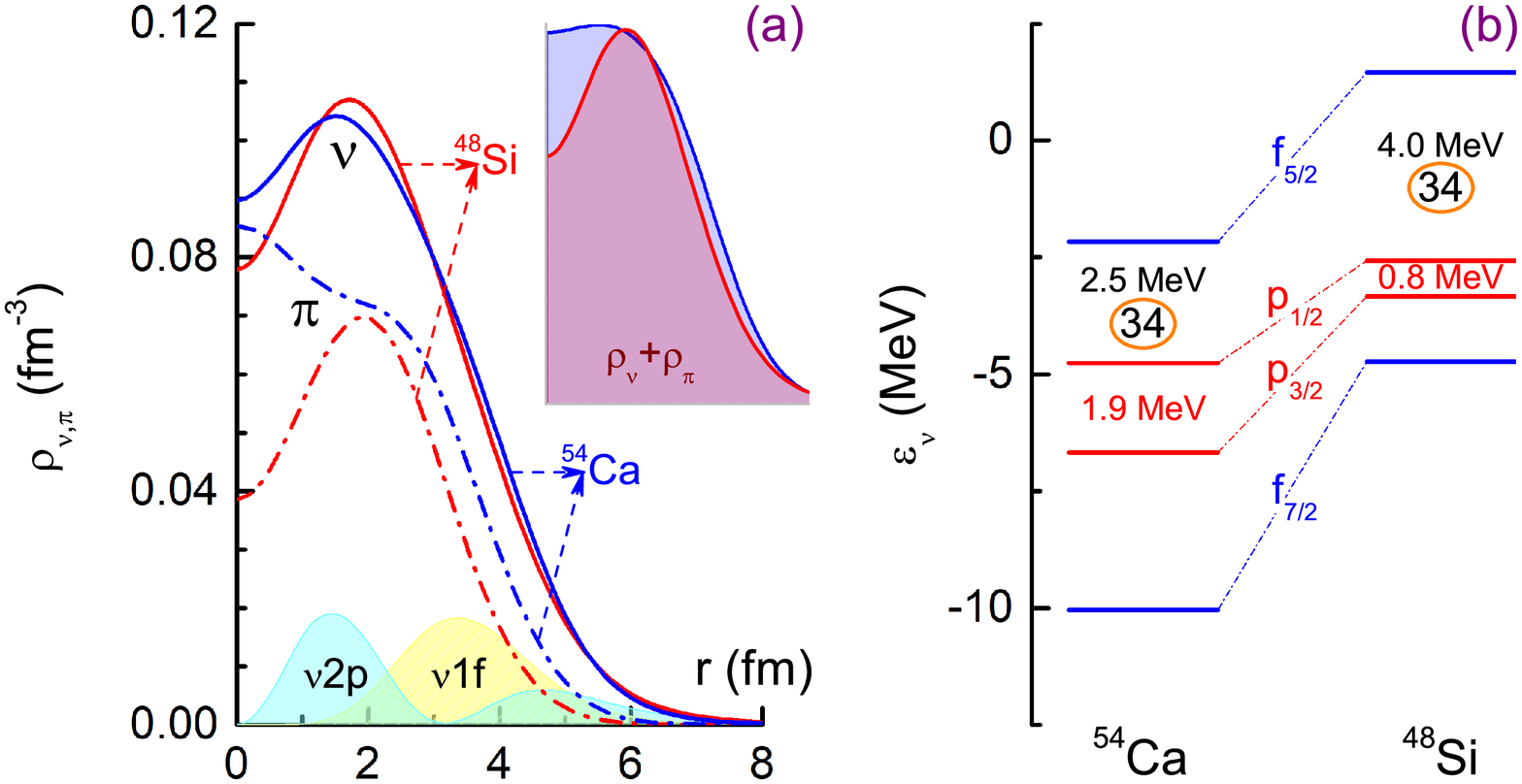}
\fi
\caption{Baryonic density distributions (a) and neutron single-particle spectra (b) for the ground states of $^{48}$Si and $^{54}$Ca, calculated by RHFB with PKA1.
The compositions of $\nu 2p$ and $\nu 1f$ for $^{48}$Si are also shown. The shell gaps of interest are indicated.}
\label{fig:CS}
\end{figure}

The neutron ($\nu$) and proton ($\pi$) density profiles for the $N=34$ isotones $^{54}$Ca and $^{48}$Si are shown in Fig.~\ref{fig:CS}(a). We observe a central depletion in both neutron densities of $^{54}$Ca and $^{48}$Si, namely the neutron bubble-like structure. There is however an important difference in their proton densities, leading to qualitatively very different total density distributions. In $^{54}$Ca the proton density exhibits a small bump in the center, which compensates the depletion of the neutron in the total density. On the contrary, more evident bubble-like structure appears in proton density profile of $^{48}$Si, which in turn makes $^{48}$Si as one of the rare but possible candidates for doubly semibubble nucleus -- with both neutron and proton bubble-like structures. In addition, a comparably large ($\sim 5.0$~MeV) proton gap $Z=14$ is found as in the doubly magic nucleus $^{34}$Si~\cite{Mutschler2017,Li2016b}. In fact, as described by ordinary mean field models~\cite{Bender2003}, $^{48}$Si has similar proton configuration as $^{34}$Si~\cite{Caurier2014,Li2016b,Mutschler2017}: the last occupied proton orbital is $\pi1d_{5/2}$, above which the $\pi2s_{1/2}$ is essentially empty. Such configurations usually occur in nuclei with a central-depressed density profile because of the lack of a $s$-state contribution~\cite{Todd2004,Khan2008,Grasso2009,Li2016b}. It is therefore a quantum shell effect, at variance with bubble-like structure in superheavy nuclei powered by the Coulomb interaction~\cite{Bender1999,Decharge1999,Li2016b,Jerabek2018}. This quantum effect can however be weakened by correlations beyond the mean field. If for instance the $s$-state is close enough to the last occupied state, pairing~\cite{Khan2008,Grasso2009,Nakada2013,Li2016b} as well as multi-reference framework beyond mean field~\cite{Yao2012,Yao2013} may populate the $s$-state with a non-zero probability, washing out the central depletion. The size of the energy gap between the last occupied state and the next $s$-state is therefore a crucial quantity to assess the prediction of a bubble-like structure. In $^{48}$Si, the existence of large proton gap ($\sim 5.0$~MeV) is a rather clear argument in favor of the occurrence of proton semibubble structure, in addition to $^{34}$Si. In fact, similar prediction is also made by the Skyrme-HFB calculation from the BRUSLIB database~\cite{Bruslib}. Notice however that while protons are predicted to have a central depletion in $^{48}$Si from the BRUSLIB database, the neutron is not. This illustrate the model dependence of such prediction and emphases its importance to better understand the nature of nuclear forces.

Let us recall that, in mean field approaches the SO interaction scales with the derivative of nuclear potential, and consistently with that of nucleon densities~\cite{Bender2003,Meng2006,Sagawa2014}.
For relativistic mean field approaches (in the first order approximation), the derivative of the isoscalar (total) density dominates the SO effects, while that of the isovector density (difference between neutron and proton densities) contributes additional but much small corrections~\cite{Reinhard1995}. Specifically, the positive radial gradient, represented by the cental depletion of nucleon density, partly compensates the negative one at the surface of nucleus, leading to a reduced SO splitting for low-$\ell$ orbitals~\cite{Todd2004,Khan2008,Grasso2009,Li2016b}. As observed in Fig.~\ref{fig:CS}(b), the $\nu2p$ splitting is reduced from $\sim1.9$~MeV in $^{54}$Ca (central depletion in the neutron density but no central depletion in the total density) down to $\sim0.8$~MeV in $^{48}$Si (doubly semibubble candidate). One may notice that the reduction of $\nu2p$ splitting ($\sim$1.1 MeV) cannot entirely account for the $N=34$ shell gap opening (increase by $\sim$1.5 MeV from $^{54}$Ca to $^{48}$Si), since the $\nu1f$ splitting is slightly reduced as well, see Fig.~\ref{fig:CS}(b). Another effect contributing to the opening of the $N=34$ shell gap is related to the properties of the pseudospin (PS) partners $\{\nu 1f_{5/2}, \nu 2p_{3/2}\}$. The spitting of these states is governed by the PSS, which is, in general, evidently broken for non-flat density distributions~\cite{Meng1998, Meng1999, Ginocchio2005, Meng2006, Liang2015}. Following the mean field analyses of Refs.~\cite{Meng1998, Meng1999}, the doubly semibubble structure will contribute to enlarge the breaking of the PPS in $^{48}$S compared to $^{54}$Ca. In conclusion, the presence of a doubly semibubble predicted in $^{48}$Si can reduce the $\nu2p$ SO splitting and enhance the splitting between the PS partners $\{\nu 1f_{5/2}, \nu 2p_{3/2}\}$, coherently triggering the emergence of a $N=34$ shell gap of $\sim4$~MeV. The doubly semibubble structure predicted in $^{48}$Si is therefore the main origin of the $N=34$ shell gap.

\begin{figure}[tb]
\centering
\ifpdf
\includegraphics[width = 0.48\textwidth]{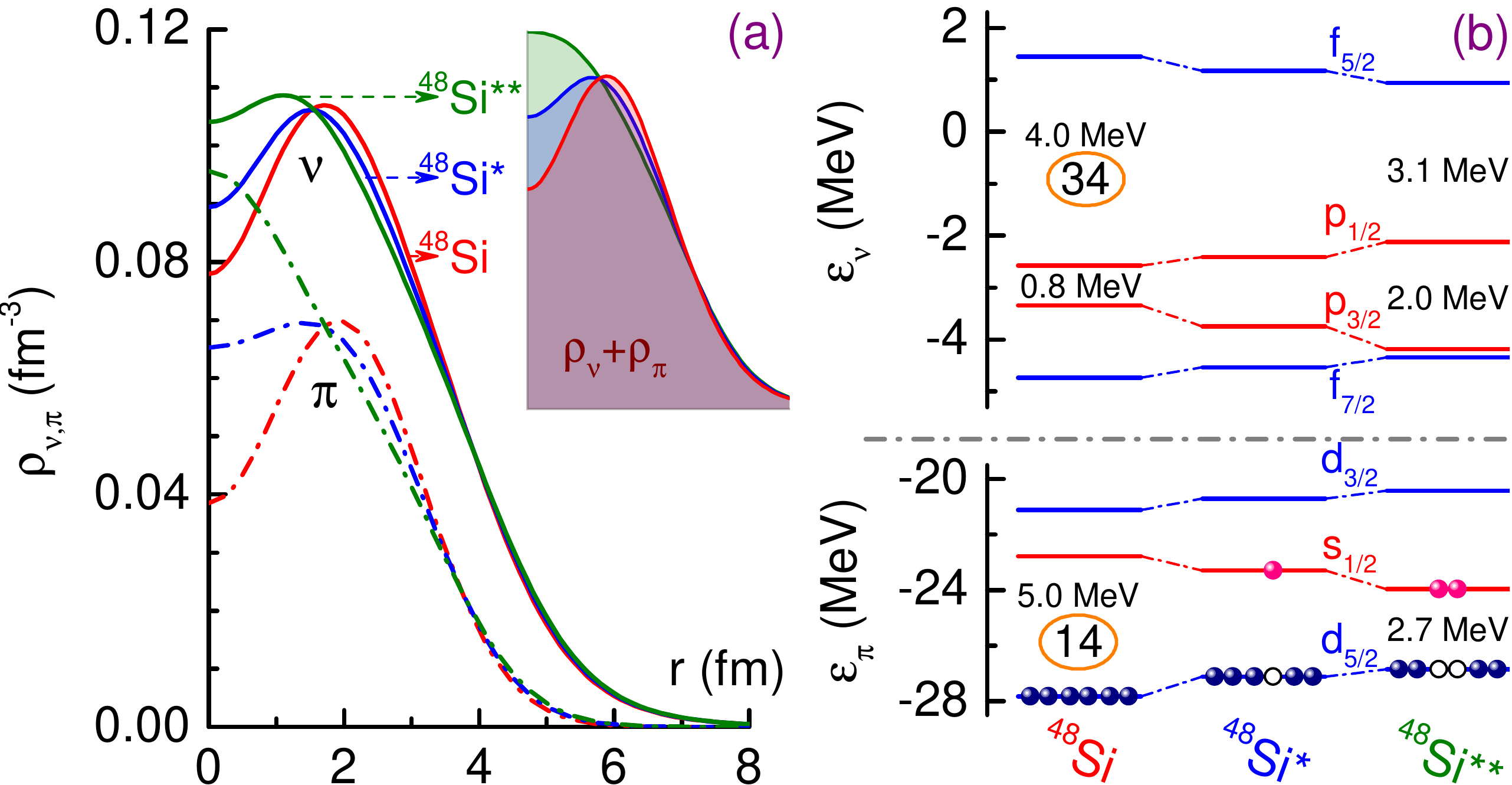}
\else
\includegraphics[width = 0.48\textwidth]{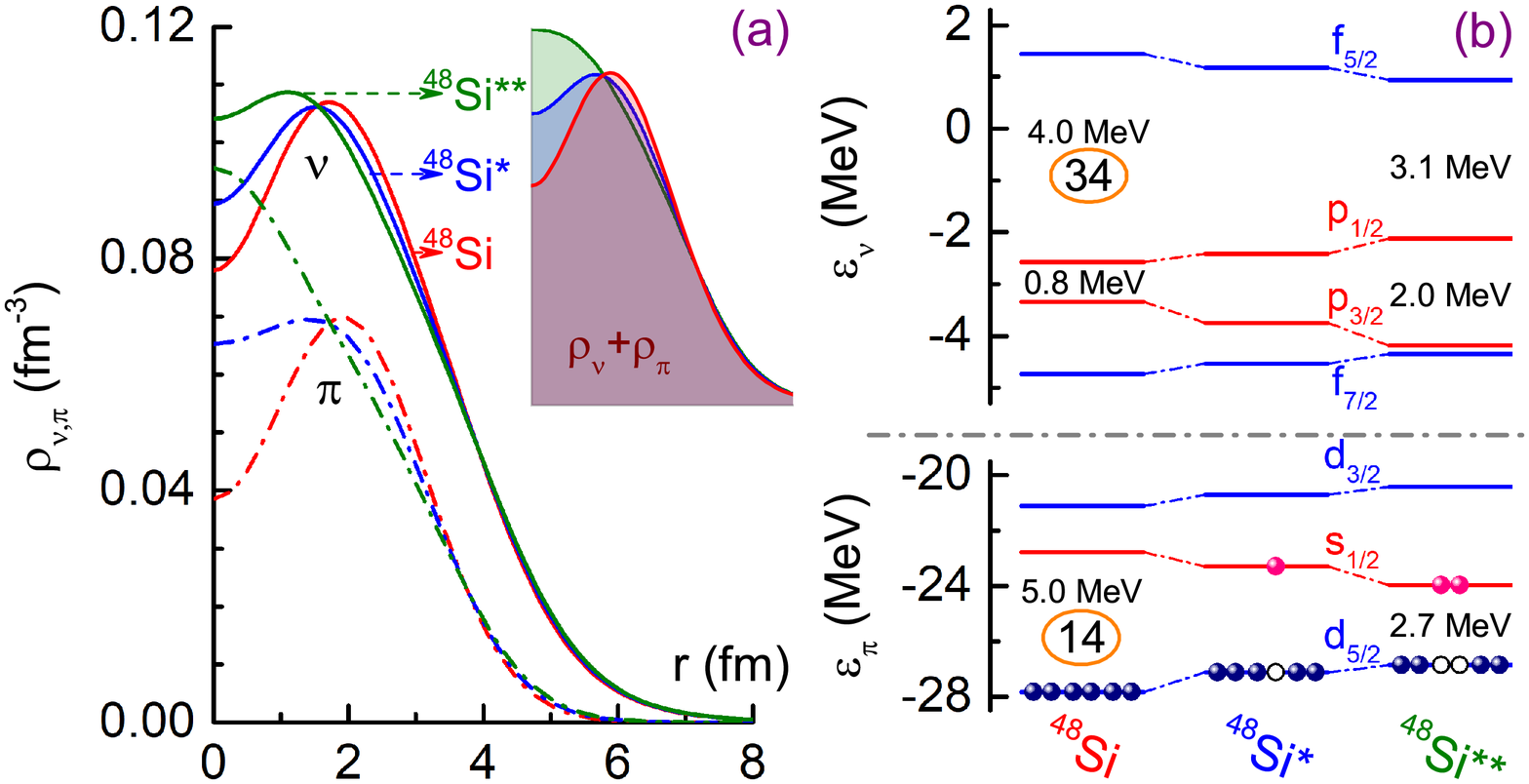}
\fi
\caption{Baryonic density distributions (a) and single-particle spectra (b) in the ground and excited configurations for $^{48}$Si,
calculated by RHFB with PKA1. Notice that the ordering of the density distributions according to their
configurations depends on the radius interval. The shell gaps of interest are indicated. See the text for details.
}
\label{fig:SE}
\end{figure}

The influence of the doubly semibubble on the $N=34$, $Z=14$ shell gaps is further analyzed in Fig.~\ref{fig:SE} where we compare the predicted ground state of $^{48}$Si with the excited states $^{48}$Si$^*$ in which $\pi 2s_{1/2}$ is partially occupied, and $^{48}$Si$^{**}$ the $\pi 2s_{1/2}$ is fully occupied. As expected, the proton central density is flatten in $^{48}$Si$^*$ and the neutron central density still manifests a depletion, reducing the effect of bubble-like structure in $^{48}$Si$^*$ compared to $^{48}$Si. While the proton central density manifests a bump in $^{48}$Si$^{**}$ and the neutron central density is flatten, washing out the central depletion in the total density, see Fig.~\ref{fig:SE}(a). There is indeed a clear relation between the sizes of $N=34$, $Z=14$ shell gaps and the central density profile: semibubble structure favors the magicities. One may also notice that the $\nu 2p$ splitting in $^{48}$Si$^{**}$ is as large as that in $^{54}$Ca, see Fig.~\ref{fig:SE}(b). This is certainly due to the fact that the total density of $^{48}$Si$^{**}$ is close to flat.

There is another interesting effect in $^{48}$Si$^*$ (or $^{48}$Si$^{**}$) for the SO splitting of the large-$\ell$ states. The increasing of the neutron(proton) central density induces a small modification of its asymptotic behaviour, as observed in Fig.~\ref{fig:SE}(a): the gradient of the peripheral nuclonic density is smaller in $^{48}$Si$^*$ (or $^{48}$Si$^{**}$) compared to $^{48}$Si. Since the $\nu 1f$ states have a larger overlap with the nuclear surface than with the interior region, contrarily to the $\nu2p$ states, the SO splitting of these larger-$\ell$ states get smaller going from $^{48}$Si towards $^{48}$Si$^*$ (or $^{48}$Si$^{**}$).  Consequently the splitting between neighboring PS partners, i.e., $\{\nu 1f_{5/2}, \nu 2p_{3/2}\}$, is somewhat enlarged. Therefore, the $N=34$ gap in $^{48}$Si$^{**}$ ($\sim 3.1$~MeV) is still larger than that in $^{54}$Ca ($\sim 2.5$~MeV). The trends discussed on neutron sector can also be seen in proton sector. With the increasing of central density, the SO splitting of $\pi1d$ is slightly reduced while the PS splitting of $\{\pi 1d_{3/2}, \pi 2s_{1/2}\}$ is enlarged. As a consequence, the $Z=14$ shell is quenched in $^{48}$Si$^{**}$, see Fig.~\ref{fig:SE}(b).

To summarize, the comparison of our predictions for the ground state of $^{48}$Si and the excited states $^{48}$Si$^*$ and $^{48}$Si$^{**}$ clearly illustrates the close relation between
i) the central depletion in the density profile and the reduction of SO splitting of low-$\ell$ orbitals, the increase of the splitting of neighboring PS doublet, and ii) the reduction of the external density profile and the decrease of the SO splitting of the larger-$\ell$ states. The mechanism for the SO reduction is the same as the one described for other nuclei~\cite{Todd2004,Li2016b} and observed in $^{34}$Si~\cite{Mutschler2017}. The PSS plays also an important role in conjunction with the onset of the doubly semibubble for the prediction for the new magic numbers $N=34$ and $Z=14$.
$^{48}$Si is therefore a special nucleus where the onset of the doubly semibubble structure induces various SO and PS couplings which coherently add together to strengthen the
new magic shells.

Let us open a parenthesis on superheavy and hyperheavy nuclei, where a similar coupling between the onset of a bubble-like (or bubble) structure and the opening of new magicities. These nuclei lie beyond the currently known region of the nuclear chart and may also take the form of bubbles~\cite{Bender1999,Decharge1999,Li2016b,Jerabek2018}. In superheavy nuclei with $Z\sim120$, the polarization due to high-$\ell$ orbitals and large Coulomb repulsion generates a central-depressed matter distribution. This makes large shell gaps possible at $Z=120$, in which the large splitting of PS doublet $\{\pi2f_{5/2}, \pi3p_{3/2}\}$ is found to coincide with the collapse of the $\pi3p$ splitting~\cite{Bender1999,Li2014,Li2016b}. Although the force producing the semibubble shape in $^{48}$Si and in $Z=120$ superheavy nuclei is different, it is interesting to stress that the magicity in these two systems may originate from the same mechanism: the presence of a bubble-like structure reduces the SO splitting and enhances the splitting between PS partners, which constructively combined together, could lead to the formation of strong shell gaps.

%
In the last part of this Letter, we now explore thermally excited states in $^{48}$Si. We remind that the coupling to the continuum, particularly the low-lying s.p. resonant states, becomes more and more important as the nucleus under study gets closer to the drip lines~\cite{Dobaczewski1996,Poschl1997,Meng2006}. For dripline nuclei the structure of the s.p. continuum, apart from its interplay with magicity, can affect the location of the drip line itself~\cite{Dobaczewski1996,Poschl1997,Pastore2013}. In $^{48}$Si we also find that the continuum has an interesting structure, where a few resonant states are located rather close to the $N=34$ shell. This structure allows a recent phenomenon to occur in $^{48}$Si, the so-called finite temperature pairing reentrance. In addition to $^{48}$Ni~\cite{Belabbas2017} and $^{176}$Sn~\cite{Margueron2012}, $^{48}$Si can be the third candidate for such phenomenon. To show this, we employ the finite-temperature RHFB approach recently developed in Ref.~\cite{Li2015}.

\begin{figure}[tb]
\centering
\ifpdf
\includegraphics[width = 0.48\textwidth]{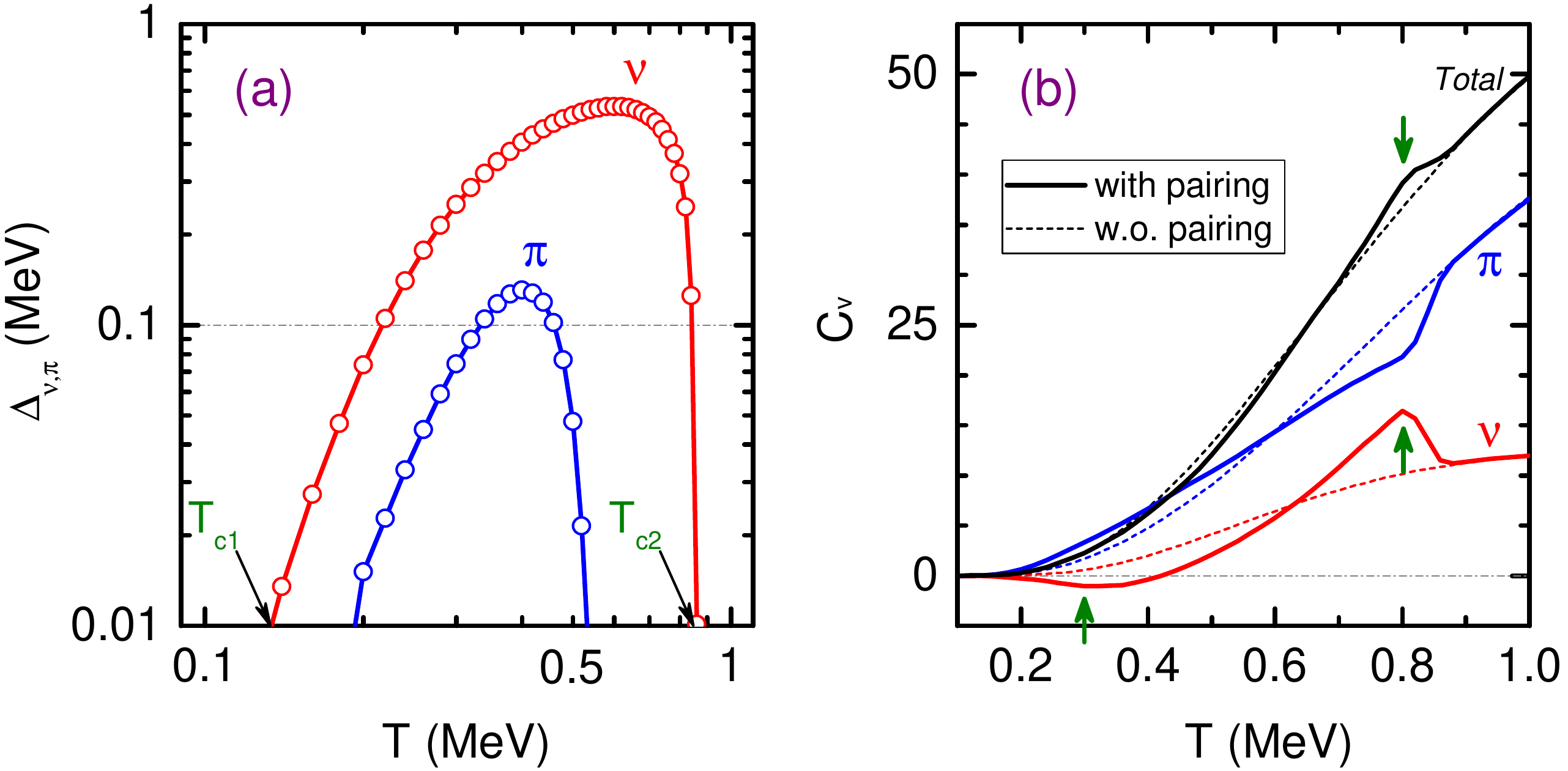}
\else
\includegraphics[width = 0.48\textwidth]{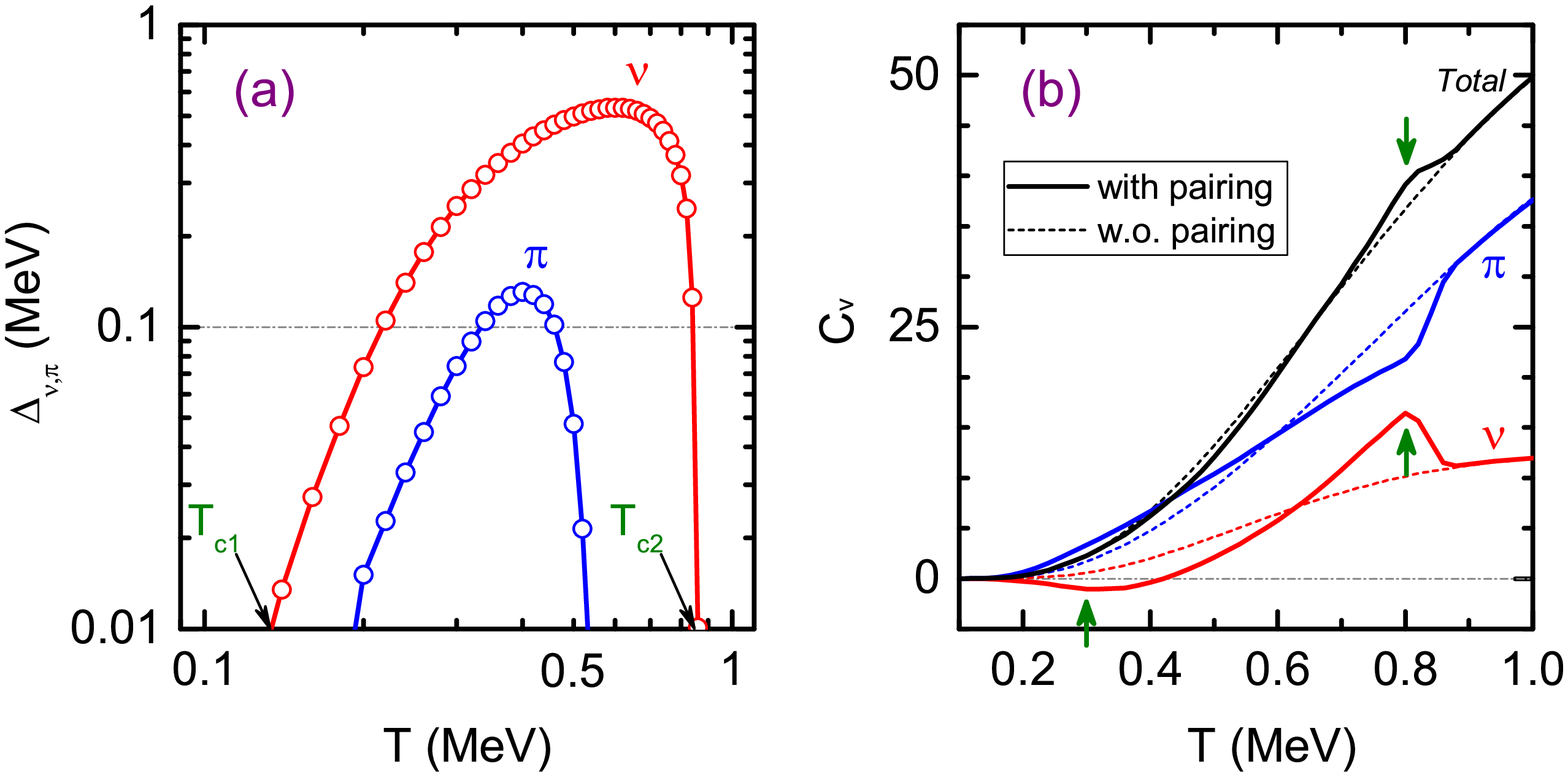}
\fi
\caption{Pairing gap (a) and specific heat (b) for $^{48}$Si as a function of temperature $T$,
calculated by finite-temperature RHFB with Lagrangian PKA1 and Gogny pairing force D1S~\cite{Berger1984}.
The signatures for phase transition are marked by arrows. See the text for details.}
\label{fig:PC}
\end{figure}

At zero and very low temperature $^{48}$Si is expected to be and to remain unpaired because of magicity. At finite temperature, however, thermal excitations may populate orbitals above Fermi level with a non-zero probability. For $^{48}$Si, the relevant states are the valence neutron orbitals $\nu2p_{1/2, 3/2}$ and $\nu1f_{7/2}$, and the continuum states above the shell $N=34$. Such a reorganisation of the orbital occupancies at finite temperature switches on pairing correlations and leads to the so-called pairing reentrance. Since this phenomenon is going against the general expectation that temperature destroys pairing, it may take place only below the usual critical temperature in finite nuclei, $\sim 1.0$~MeV~\cite{Margueron2012,Niuyf2013,Li2015}. In Fig.~\ref{fig:PC}(a), we represent the neutron pairing gap $\Delta_\nu$ for $^{48}$Si as a function of temperature. There are two critical temperatures corresponding to the low- and high-temperature boundaries of pairing reentrance, which are predicted to be $T_{c1}\sim0.15$ MeV and $T_{c2}\sim0.90$~MeV. Outside this temperature interval, neutrons in $^{48}$Si are in the normal (unpaired) phase.
It is worth noticing that protons also manifest pairing reentrance which appears from $T=0.2$~MeV and quenches at $T=0.5$~MeV, but the effects are negligible. This is because i) the $Z=14$ proton shell is robuster than the $N=34$ neutron shell and ii) the relevant valence states around the Fermi level in proton sector are much less than that of neutron's.

The specific heat $C_V$ reflects the second derivative of the free energy with respect to the temperature and it is thus sensitive to the thermal excitations of nucleus. Figure~\ref{fig:PC}(b) shows the comparison between the specific heats obtained by switching on and off the pairing correlations for $^{48}$Si. At low temperature, the specific heat of the normal phase (in dashed lines) is almost a linear function of the temperature, whereas that of the neutron superfluid phase (in solid lines) is strongly suppressed at low temperature, above $T_{c1}$, as expected (see the lowest arrow).
At around $T = 0.8$~MeV, these is a bump in the total specific heat, originated from the neutron side (see the second). At higher temperature ($\sim 1.0$) pairing vanishes. The peculiar structure of the specific heat between $T_{c1}$ and $T_{c2}$, see Fig.~\ref{fig:PC}(a), can be considered as a signature of pairing reentrance. Moreover, although protons present rather weak pairing, their specific heat is different from the normal phase, showing that protons are also sensitive to the phase transition occurring in the neutron channel since they are cross-interacting. We find that the proton and neutron specific heats vary with opposite behaviors between $T_{c1}$ and $T_{c2}$, thus reducing the impact of the neutron superfluidity in the total specific heat. We therefore predict a weak but still visible bump in the total specific heat at $T\sim 0.8$~MeV.

In our framework, particle number is imposed only on average from the chemical potential. It has been shown within the Bardeen-Cooper-Schrieffer (BCS) approximation including particle number restoration that, in some cases, doubly magic nuclei could be weakly paired even in their ground state~\cite{Hung2008,Gambacurta2013}. This is an interesting additional effect which is not included in our framework, but this prediction does not go against our prediction at finite temperature, since these two kinds of correlations --- particle number restoration and finite temperature --- both act towards the same direction: enhancement of pairing correlations. In the future, it will however be interesting to perform a finite-temperature RHFB calculation with particle number restoration.

The confirmation of pairing reentrance in $^{48}$Si represents very challenging predictions for both nuclear physics rare-isotope beam facilities and theoretical developments at the frontier of stability. Signatures of pairing reentrance may be low-temperature anomalies of the specific heat or of the level density, or even might be deduced from pair transfer reaction mechanism~\cite{Shimizu1989,Dean2003}. Interestingly, a similar phenomenon called pairing persistence~\cite{Margueron2012,Li2015} was recently predicted from finite temperature approaches and may occur in less exotic nuclei, close to subshell closure. The pairing persistence leads to an increase of the critical temperature beyond the usual BCS limit, powered by thermally excited states around the Fermi level. At low temperature (below $~1.0$~MeV) pairing can persist beyond the BCS limit due to the structure of the excited states. The mechanism for pairing persistence and pairing reentrance is therefore similar in nature, while occurring in different systems.

%
In summary, based on the self-consistent relativistic mean field approach, we predict that $^{48}$Si may be the next {\it doubly magic drip-line} nucleus. Such prediction is made by employing a most complete RHFB Lagrangian PKA1. Our prediction is model dependent and other Lagrangians, such as the PKOi series does not lead to the same prediction. The PKOi are however less complete in the kind of interacting vertex than PKA1 and have already proven difficulties in predicting recent new magicities in neutron rich nuclei that PKA1 could reproduce~\cite{Li2016a}. Independently of the model themselves, one can remark that $^{48}$Si is at the intersection between new magicities observed for neutron and proton, i.e., $Z=14$ in neutron rich light nuclei~\cite{Burgunder2014, Mutschler2017} and $N=34$ expected from measurements in $^{54}$Ca~\cite{ Steppenbeck2013,Michimasa2018}. $^{48}$Si is also predicted to be the first candidate of {\it doubly semibubble} nucleus with both neutron and proton bubble-like shapes. A novel mechanism for opening of new magicities is illustrated in this Letter, where the doubly bubble-like structure produces a reduction of the SO splitting and an increase of the PS splitting, which coherently lead to the $N=34$, $Z=14$ shell gaps in $^{48}$Si. We suggest that measurements of $E(2^+_1)$ and $B(E2)$ values in isotones close to $^{48}$Si, such as $^{50}$S for instance, will already provide trends which could be compared to theoretical predictions. It is appealing to notice that new neutron-rich nuclei $^{47}$P, $^{49}$S that close to drip lines have been discovered very recently~\cite{Tarasov2018}.

Moreover, $^{48}$Si is potentially one of the very few nuclei for which {\it pairing reentrance} occurs at finite temperature. The occurrence of such phenomenon depends sensitively on the structure of the s.p. spectrum, like the size of the shell gap and the type of resonant states, which impacts the values for the critical temperatures $T_{c1}$ and $T_{c2}$, as well as the temperature dependence of the specific heat. There is therefore a very strong model dependence in our prediction and we have shown that $^{48}$Si may be a very atypical nucleus benchmarking theoretical modeling. The weakly bound $^{48}$Si nucleus is only four neutrons beyond the heaviest isotope presently observed, $^{44}$Si, making it yet unknown but probably accessible for the next generation of radioactive ion beam facilities. The confirmation or the refutation of our predictions for the ground state and excited states of $^{48}$Si represents therefore both an experimental and theoretical challenge for the understanding of the nature of nuclear forces.

\section*{Acknowledgements}

This work is partly supported by the National Natural Science Foundation of China under Grant No. 11675065.
J.L. acknowledges the support by the Alexander von Humboldt foundation.


%
%
\end{document}